\documentclass[aps, prb, reprint, showkeys, nolongbibliography]{revtex4-2}

\pdfoutput=1

\usepackage[utf8]{inputenc}
\usepackage[usenames,dvipsnames]{color}
\usepackage{hyperref}
\usepackage{amssymb}
\usepackage{amsmath}
\usepackage{multirow}
\usepackage{epsfig}
\usepackage{rawfonts}
\usepackage{latexsym}
\usepackage{theorem}
\usepackage{natbib}
\usepackage{graphicx}

\begin{document}

\title{Quantum Monte Carlo study of artificial triangular graphene quantum dots}
\date{\today}

\author{E. Bulut Kul$^1$, G\"{o}khan \"{O}ztarhan$^1$,
  M. N. Çınar$^2$, and A. D. G\"{u}\c{c}l\"{u}$^1$}
\affiliation{$^1$Department of Physics, \.{I}zmir Institute of Technology, 35430, \.{I}zmir, Türkiye \\
  $^2$Department of Materials Science and Engineering, \.{I}zmir
  Institute of Technology, 35430, \.{I}zmir, Türkiye}

\begin{abstract}

  We investigate the magnetic phases of semiconductor-based artificial
  triangular graphene quantum dots (TGQDs) with zigzag edges using
  variational and diffusion Monte Carlo methods. These systems serve
  as quantum simulators for bipartite lattices with broken sublattice
  symmetry, providing a platform to study the extended Hubbard model’s
  emergent magnetic phenomena, including Lieb’s magnetism at
  half-filling, edge depolarization upon single-electron addition, and
  Nagaoka ferromagnetism. Our nonperturbative quantum Monte Carlo
  simulations, performed for finite-sized TGQDs modeled as
  nanopatterned GaAs quantum wells, with system sizes up to $N_s=61$
  lattice sites, reveal a transition from metallic to insulating
  regimes as a function of the quantum well radius $\rho$, while
  preserving edge-polarized ground states at half-filling. Notably,
  edge depolarization occurs upon single-electron doping in both
  metallic and insulating phases, in contrast to the Nagaoka
  ferromagnetism observed in hexagonal armchair geometries.

\end{abstract}

\keywords{artificial graphene, graphene quantum dots, quantum
  simulators, variational Monte Carlo, diffusion Monte Carlo}

\maketitle
Recent advances in the fabrication of artificial superlattices have
enabled reliable quantum simulators, allowing known physical
phenomena to be reproduced at accessible length and energy scales,
while also enabling the exploration of new physics
\cite{Bloch2005,Mazurenko2017,Weimer2010,Islam2011,PhysRevX.8.031022,Buluta2009,nww023,Salfi2016,Aspuru-Guzik2012,Cai2013,Bernien2017,li21,Guo2024}.
Among these superlattices, artificial graphene nanostructures,
realized with cold atoms
\cite{Soltan-Panahi2011,Tarruell2012,Uehlinger2013} and
semiconductor heterojunctions
\cite{Simoni2010,Du2018,AG_observation_of_dirac_bands},
have proven effective for studying massless Dirac fermions. Recent
observations of graphene-like behavior in nanopatterned AlGaAs/GaAs
quantum wells are particularly advantageous due to the ease of
fabrication and precise tunability of parameters, such as the Hubbard
ratio $U/t$
\cite{Du2018,AG_observation_of_dirac_bands}. These developments have
prompted theoretical investigations into artificial graphene quantum
dots of specific sizes and shapes, including triangular quantum dots
with zigzag edges
\cite{Yasser2022,Oztarhan2023,Ortiz_2023,ABDELSALAM2024116059,Madail2024,Saleem2024,Garcia-Ruiz2024}.

Triangular zigzag graphene quantum dots (TGQDs) have attracted
considerable attention due to their unique electronic and magnetic
properties \cite{WangWei2007,Rossier2007,Heiskanen2008,Ezawa2008,Potasz2011,graphene_devrim_book,Potasz2012,Hagymasi2018,Han2024}.
Unlike other GQDs, the distinct shape and edge structure of
triangular zigzag-edged GQDs give rise to localized edge states that
exhibit magnetic moments consistent with Lieb's theorem, making them
particularly promising for spintronic
applications \cite{Senger2010,Sun2017,Gregersen2017}. Furthermore,
the triangular zigzag edges enable tunable optical and electronic
properties through size control, rendering them versatile for
applications in quantum computing, photovoltaics, and
catalysis \cite{Guclu2013,Li2015,Wang2016,Yassine2024}. The
significance of GQDs extends beyond technological potential, as they
also provide valuable insights into fundamental quantum mechanical
behaviors in low-dimensional systems
\cite{Christopher2018,Li2021,Hong2024}.

A particularly interesting property of real TGQDs is the edge
depolarization effect, which is predicted by configuration interaction
calculations performed within the subspace of degenerate edge orbitals.
When an extra electron is added to the edge-polarized, charge-neutral
system, edge magnetization is lost \cite{Devrim2009,Potasz2012},
enabling optical control of magnetic and transport
properties \cite{Guclu2013}. This depolarization effect was recently
investigated in semiconductor-based artificial TGQDs using a
configuration interaction approach restricted to the degenerate edge
band \cite{Yasser2022}. It was found that, while the ground-state spin
does not fully depolarize upon electron addition, the relatively large
gap separating the spin-polarized ground state from other states
collapses dramatically, rendering the edge polarization unstable.

To the best of our knowledge, no methodology other than
configuration interaction has been used to confirm the collapse of edge
magnetism in real or artificial TGQDs, presumably because the
depolarization is a strong correlation effect that does not appear in
mean-field approximation-based calculations. In semiconductor-based
artificial TGQDs, the Hubbard parameter $U/t$ can reach values of the
order of 100 \cite{Oztarhan2023}, i.e., two orders of magnitude larger
than in real TGQDs. Consequently, the energy gap protecting the edge
states from the bulk can become significantly smaller than the
electron-electron interaction scale, casting doubt on the accuracy of
perturbative approaches. Moreover, recent work has shown that the
addition of a single electron can induce Nagaoka ferromagnetism in
hexagonal armchair-type artificial graphene
nanostructures \cite{Oztarhan2025}. Whether this mechanism competes
with edge depolarization remains an open question. Furthermore, the
applicability of Lieb's theorem for the Hubbard $U/t$ model to
semiconductor artificial structures is not immediately clear,
particularly in cases where long-range interactions and the
next-nearest-neighbor tight-binding (TB) hopping term $t'$ may play a
significant role \cite{Oztarhan2023}. In this work, we employ
continuum quantum Monte Carlo (QMC) methods, including variational and
diffusion Monte Carlo, where, in contrast to discrete lattice QMC,
electrons are confined in continuum potentials and electron-electron
interactions are treated non-perturbatively. This approach allows us to
accurately investigate the applicability of Lieb’s theorem at
half-filling, the emergence of edge depolarization effects, and the
stability of Nagaoka ferromagnetism in charged artificial TGQDs of
varying system sizes, characterized by the number of lattice sites
($N_s=33,\,46,\,61$) in the confining potentials.

\begin{figure}[h]
  \centering
  \includegraphics[width=0.950\columnwidth]{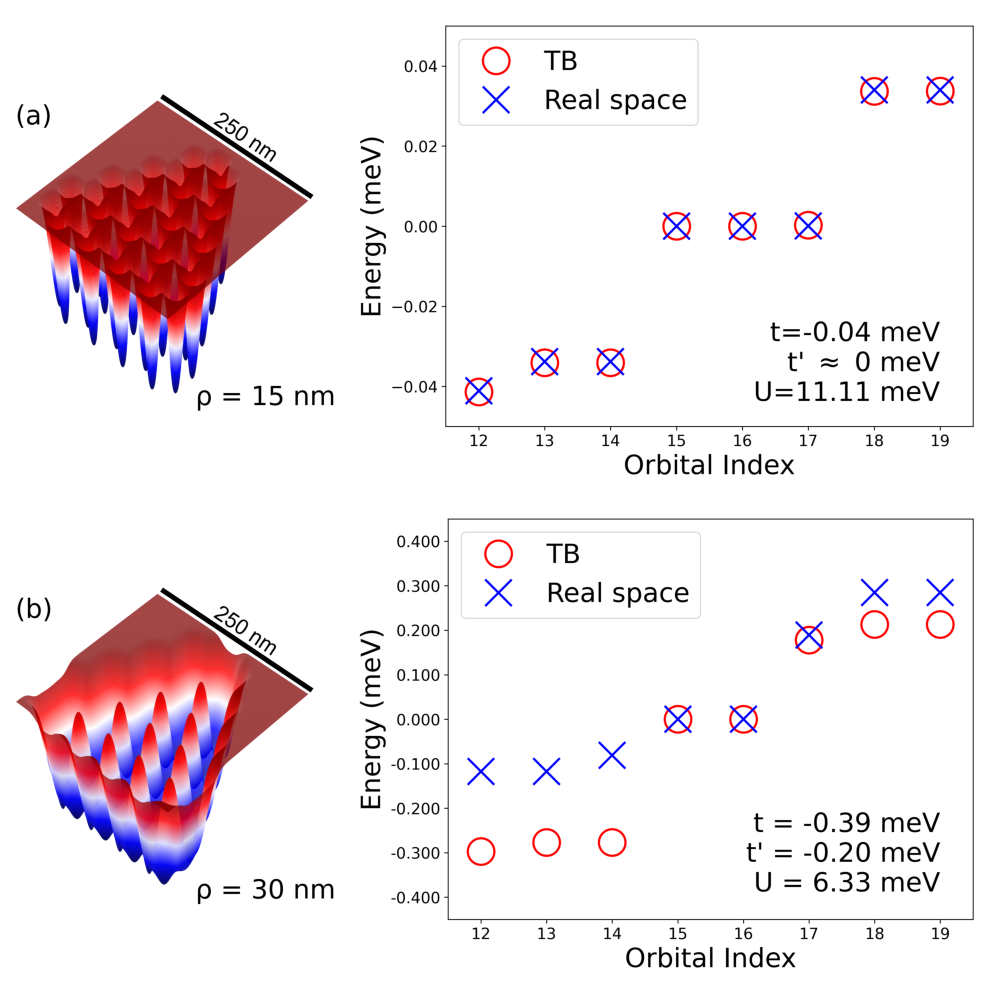}
  \caption{\label{Fig1}
    Potential profiles (left panels) and single-particle spectra
    (right panels) for a 33-site TGQD constructed using Gaussian-like
    potentials with radius (a) $\rho=15$ nm and (b) $\rho=30$
    nm. Energy spectra are obtained using the real-space finite
    differences method and compared to TB solutions with fitted
    parameters. The total lateral size of the structure is
    approximately 250 nm.}
\end{figure}
The system of $N$ interacting electrons in a honeycomb array of $N_s$
confining potentials is described by the many-body Hamiltonian
\begin{align} \label{hamiltonian}
  H = -\frac{1}{2} \sum_{i=1}^N \nabla^{2}_{i} + \sum_{i=1}^N V(\textbf{r}_{i})
      + \sum_{i=1}^N k |\textbf{r}_{i}|^2 + \sum_{i<j}^N \frac{1}{r_{ij}},
\end{align}
in effective atomic units (electronic charge $e$, dielectric constant
$\epsilon$, effective mass $m^*$, and $\hbar$ are set to 1), where $k$
is the spring constant of a quadratic gate potential at the center of
the system. This term controls finite-size effects and helps preserve
charge uniformity; without it, electrons tend to double-occupy edge
sites \cite{Oztarhan2023}. $V(\textbf{r}_{i})$ is the total potential
energy of the confining potentials. Typical material parameters for
GaAs are used, with effective electron mass $m^{*} = 0.067 m_{0}$ and
dielectric constant $\epsilon = 12.4$, yielding an effective Bohr
radius $a^{*}_{0} = 9.794$ nm and effective Hartree energy of 11.857
meV. The honeycomb array of potential wells is modeled using
Gaussian-like functions \cite{AG_theoretical_3}:
\begin{align} \label{potential}
    V(\textbf{r}) = V_{0} \sum_{k=1}^{N_s}
    \exp\Big[-\Big(\frac{|\textbf{r}-\textbf{R}_{k}|^{2}}{\rho^{2}}\Big)^{s}\Big],
\end{align}
where $V_{0}$ is the potential depth, $\rho$ is the radius, and $s$
controls the sharpness of the wells located at $\textbf{R}_{k}$. In
our calculations, the nearest-neighbor distance and Gaussian sharpness
are fixed at $a=50$ nm and $s = 1.4$, respectively. The values of
$V_0$ vary depending on the quantum well radius and system size, chosen to
satisfy the charge-neutrality condition as in our previous work
\cite{Oztarhan2023}. A 3D visualization of 33 artificial TGQD sites
constructed using Gaussian-like potentials is shown in Fig. 1 for
$\rho=15$ nm and $\rho=30$ nm.

In our calculations, trial wave functions were first obtained using
tight-binding (TB) and self-consistent mean-field Hubbard (MFH)
single-particle orbitals, directly modeling finite-sized TGQDs with
$N_s=33,\,46,\,61$ lattice sites. The Slater-Jastrow trial wave
functions were then constructed from these orbitals, each consisting
of a single Slater determinant, with the Jastrow factor included as
described in Ref.~\cite{qmc_devrim_0}. While TB-based trial wave
functions are well suited to describe metallic phases, the adjustable
$U_t/t$ ratio of the MFH method allows for the generation of both
liquid-like ($U_t/t=2$) and localized ($U_t/t=20$) trial wave functions.
The trial wave functions were optimized using the VMC method and then
used as input for the fixed-node diffusion Monte Carlo (DMC)
calculations. This approach allows us to obtain a more accurate
upper bound to the many-electron ground state of artificial TGQDs for
neutral and gated systems. Variational and fixed-node energies of
these three types of orbitals are expected to indicate a possible
transition from a metallic state to an antiferromagnetic order as a
function of $\rho$. In this work, all quantities that do not commute
with the Hamiltonian were calculated using an extrapolated estimator,
$\langle \hat{O} \rangle = 2\langle \hat{O} \rangle_{DMC} - \langle
\hat{O} \rangle_{VMC}$ \cite{qmc_review}. In DMC calculations, the
time-step was fixed at $\tau=0.05$ after testing for time-step errors.
We employed a target number of walkers of $\sim 50$ per thread and
110 threads to keep population control error below the statistical
error.

Before investigating the many-body effects, we study single-particle
properties computed using the finite differences (FD) method on a
1001$\times$1001 grid for a 33-site TGQD shown in
Fig. 1. TGQDs with zigzag edges and $N_s$ sites exhibit $N_{deg}$
degenerate eigenstates in the middle of the TB-predicted
electron-hole symmetric energy spectrum, such that
$N_s = N_{deg}^2 + 6N_{deg} + 6$ \cite{Ezawa2007,graphene_devrim_book,PotaszG2010}. For $N_s=33$,
there are three degenerate states, corresponding to eigenstate
indices 15, 16, and 17 shown in Fig. 1. A fit of the FD spectrum to
the TB spectrum within the second nearest neighbor approximation is
then performed by optimizing the nearest-neighbor hopping $t$ and
second-nearest-neighbor hopping $t'$ to minimize the difference
between FD and TB eigenenergies. For $\rho=15$ nm, the FD-TB fitting
reveals $t=-0.04$ meV, while $t'$ is negligibly small. An estimation
of the Hubbard parameter for a single site using
$U = 2 \pi \int r n(r) V_{e}(r) dr$ (verified through our QMC
results), different from the trial wave function Hubbard $U_t$, gives 11.11 meV, approximately 200 times larger than $t$.
For $\rho=30$ nm, the value of the second-nearest-neighbor hopping
$t'$ approaches $t$, and the TB approach no longer provides a good
approximation; a satisfactory fit between FD and second nearest neighbor TB cannot be obtained.
More importantly, the edge states with orbital indices 15, 16, and 17
are no longer protected by a gap, and their degeneracy is broken. We
also note that the confinement potential shape for large $\rho$ mimics
the honeycomb lattice obtained using a triangular lattice of
antidots \cite{Du2018}, as seen in the left panel of Fig. 1b. In the
following, we investigate many-body ground states for various values
of $\rho$ between 15–35 nm, for both half-filled and charged systems.

\begin{figure}[h]
\centering
\includegraphics[width=1.0\columnwidth]{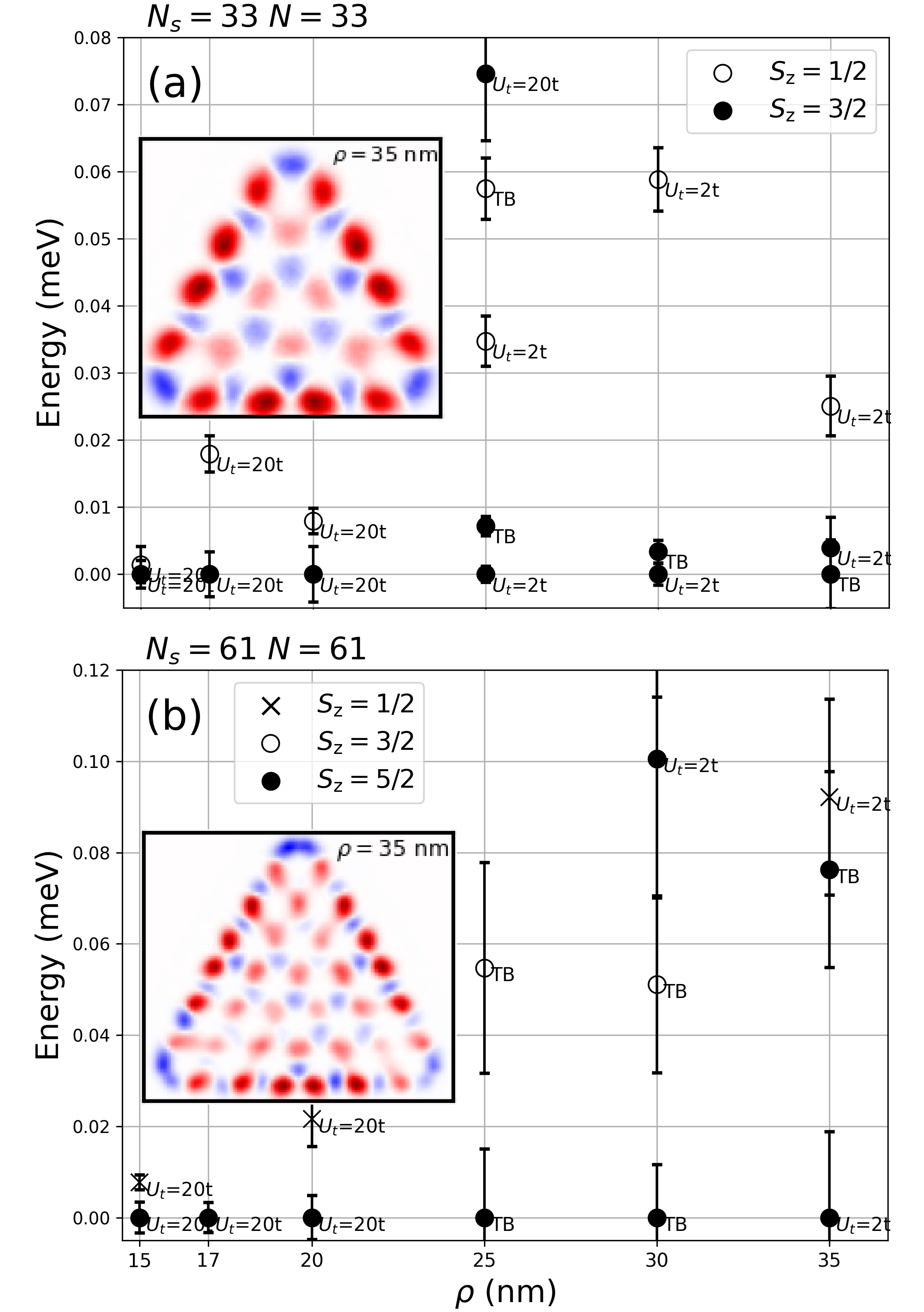}
\caption{\label{Fig2} Diffusion Monte Carlo energies for
  charge-neutral TGQDs as a function of the quantum well radius
  $\rho$ for (a) $N_s=33$ sites and (b) $N_s=61$ sites, for all
  possible spin excitations within edge states, obtained using
  various trial wave functions. Insets (a) and (b) show the lowest
  energy spin densities for $\rho = 35$.}
\end{figure}
In Fig. 2a, we investigate the ground state total spin $S_z$ for
half-filled (charge neutral) system with $N_s=33$ sites as a function
$\rho$. The three degenerate states discussed above host 3 electrons
which can have total spin $S_z=1/2$ or $3/2$.  According to Lieb's
theorem which applies to Hubbard model with no long-range
interactions, total spin $S$ must be half of the difference between the
number of atoms in sublattices $A$ and $B$,
i.e. $S=(N_A-N_B)/2=3/2$. Our single-determinant QMC trial wave
functions that were built in the subspaces of $z$-projection of spin
using Hubbard and TB orbitals systematically leads to ground states
with $S_z=3/2$ in agreement with Lieb's theorem, except for $\rho=15$
where the ground state could not be determined due
to statistical error. Moreover, for lower values of $\rho$, the $U_t=20t$
trial wave function gives a better fixed-node energy, whereas for
larger $\rho$ values, $U_t=2t$ or TB based trial wave functions are
better. Similar results were obtained for a $N_s=61$ sites TGQD, shown
in Fig. 2b. Since the system now exhibits 5 degenerate edge states, we
scanned the subspaces $S_z=1/2$, $3/2$, and $5/2$ using Hubbard and TB
trial wave functions, and found that edge states are polarized, giving
$S_z=5/2$ ground state for all $\rho$ values.

\begin{figure}[h]
  \centering
  \includegraphics[width=1\columnwidth]{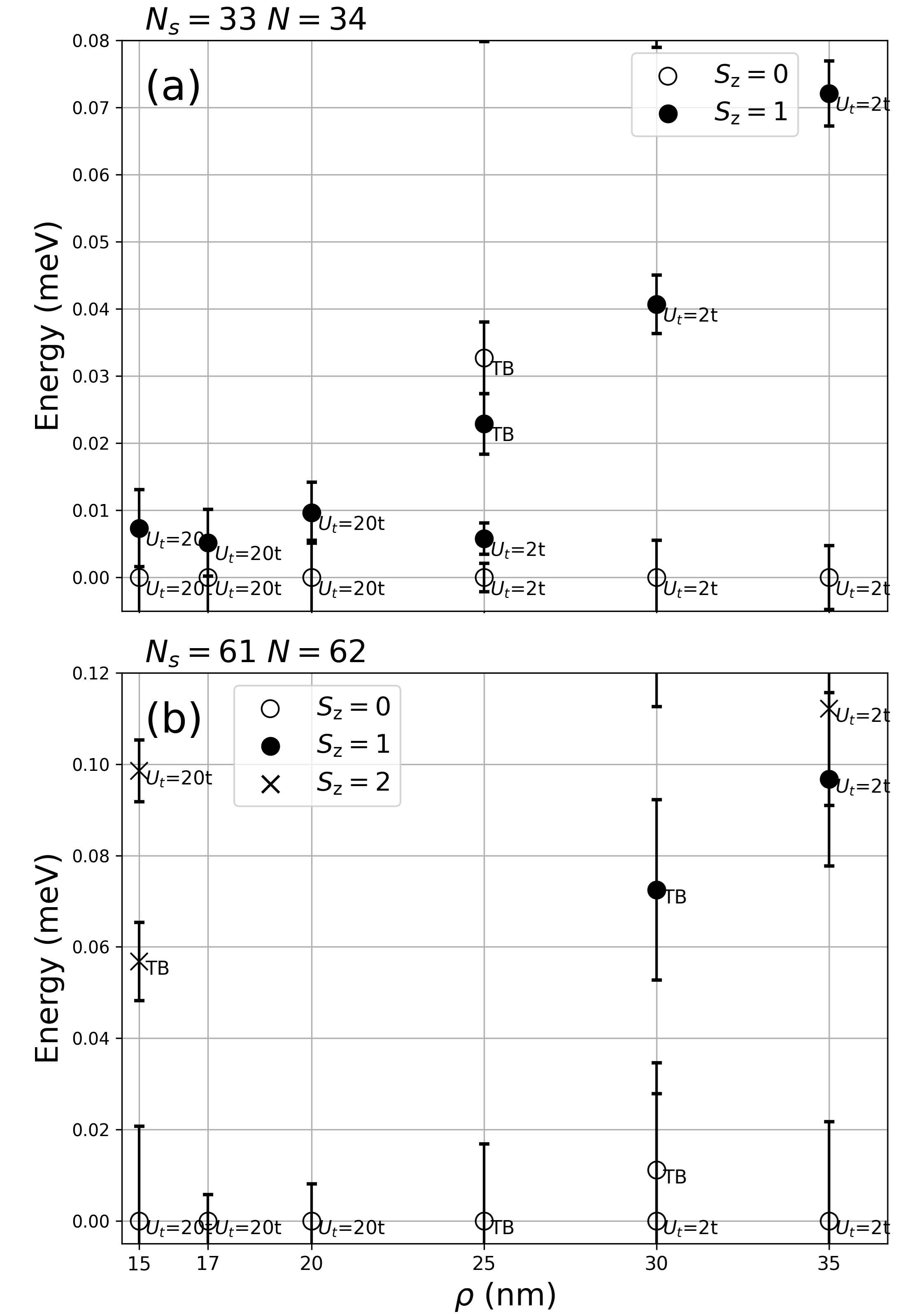}
  \caption{\label{Fig3} Diffusion Monte Carlo energies for TGQDs with
    an additional electron as a function of the quantum well radius
    $\rho$ for (a) $N_s=33$ sites and (b) $N_s=61$ sites, for all
    possible spin excitations within edge states, obtained using
    various trial wave functions.}
\end{figure}
Next, we focus on whether the spin depolarization effect predicted for
real TGQD structures is also present in artificial TGQDs. This effect
manifests as the minimum total spin ($S$) when an extra electron is
added to the neutral system \cite{Guclu2009}. A TGQD with 33 sites and
34 electrons has two competing spin configurations, $S_z = 1$ and
$S_z = 0$. Our numerical results show that, while the energy
differences are small, depolarization occurs for $\rho$ values
between 15 and 35 nm, as seen in Fig. 3a. For the larger structure
with 61 sites and 62 electrons (Fig. 3b), the depolarization effect
is even stronger, with a larger gap separating the $S_z = 1$ and
$S_z = 0$ configurations.

\begin{figure}[h]
  \centering
  \includegraphics[width=1\columnwidth]{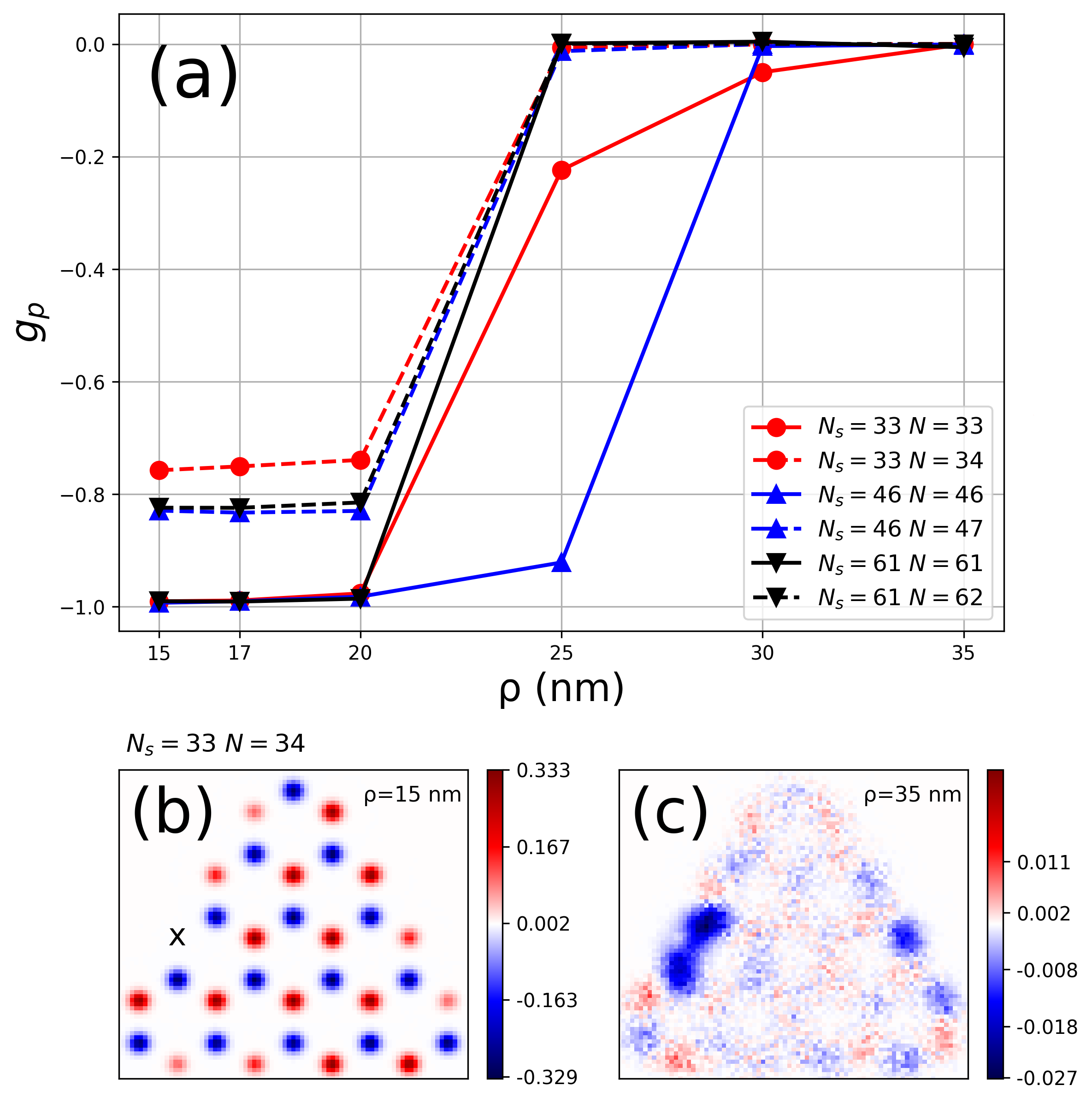}
  \caption{\label{Fig4} (a) Extrapolated spin-spin correlation
    functions $g_p$ obtained via pair densities for neutral and gated
    TGQDs with $N_s=33, 46$, and $61$ sites. Corresponding statistical
    error bars are smaller than the size of the symbols. (b,c)
    Extrapolated pair-density results for $N_s = 33$ and $N = 34$,
    corresponding to the ground state at $\rho = 15\,\mathrm{nm}$ and
    $35\,\mathrm{nm}$. The fixed reference electron position is marked
    by \textbf{x}.}
\end{figure}
In Fig. 4, we investigate the internal spin structure of charge-neutral
and charged systems with $N_s=33, 46,$ and $61$ sites. In our previous
work \cite{Oztarhan2023}, we showed that charge-neutral hexagonal
armchair and triangular zigzag artificial graphene dots undergo a
metallic-to-AFM phase transition around $\rho=20$ nm as the radius is
decreased. To investigate whether a similar transition occurs for gated
systems, we consider a real-space spin-spin correlation function
averaged over all nearest neighbors $(i,j)$, defined as
$g_p=\langle m_i m_j \rangle / \langle n_i n_j \rangle$, where $m_i$
and $n_i$ are the average magnetization and electron density at site
$i$ within a radius $r=a/2$. To reveal the internal spin structure,
$m_i$ and $n_i$ are calculated from pair densities
$p_{\sigma \sigma_0}(r, r_0)$, which represent the probability of
finding an electron with spin $\sigma$ at location $r$ when an
electron with spin $\sigma_0$ is fixed at location $r_0$. The spin-spin
correlation function values remain in the range $[-1, 1]$, with
$g_p=-1$ corresponding to AFM and $g_p=0$ corresponding to the metallic
configuration. Interestingly, Fig. 4a shows that a metallic-to-AFM
transition also occurs when the system is charged with an extra
electron (dashed lines). At lower values of $\rho$, $g_p$ is close to
$-1$. Although perfect AFM order is not possible when an extra electron
is present, $g_p$ reaches approximately $-0.8$, indicating relatively
strong AFM behavior. This is visualized in Figs. 4b and 4c, showing
the pair spin density
$p_{\uparrow \uparrow}(r, r_0)-p_{\downarrow \uparrow}(r, r_0)$,
where the reference spin-up electron fixed on top of a site is shown
with a cross. For $\rho=15$ nm, AFM configuration is clearly visible,
whereas for $\rho=35$ nm only short-range spin-spin correlations are
observed near the fixed electron, indicating metallic behavior.

\begin{figure}[h]
  \centering
  \includegraphics[width=1\columnwidth]{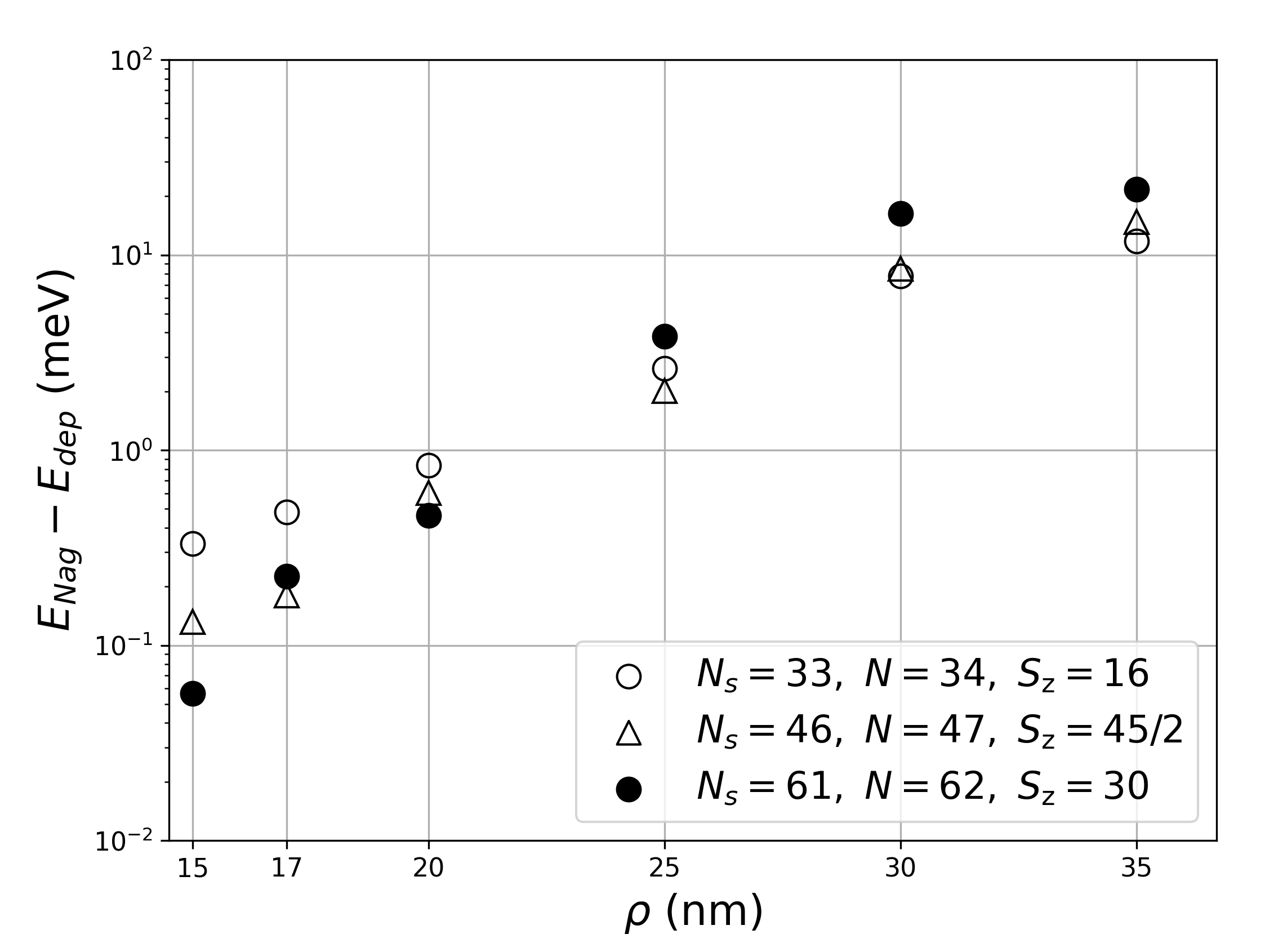}
  \caption{\label{Fig5} Diffusion Monte Carlo energy difference
    between the Nagaoka ferromagnetic state and the edge-depolarized
    state, $E_{Nag}-E_{dep}$, for $N_s=33, 46,$ and $61$ site TGQDs,
    showing that Nagaoka ferromagnetism does not occur as the ground
    state. Statistical error bars are smaller than the size of the
    symbols.}
\end{figure}

Another possible magnetic phase transition induced by the addition of
a single charge in graphene quantum dots is Nagaoka
ferromagnetism \cite{Nagaoka1966,Oztarhan2025}. Initially predicted in
the infinite on-site repulsion limit ($U \to \infty$) of the strongly
correlated Hubbard model, Nagaoka ferromagnetism is expected to occur
in the presence of a hole (or an electron, if electron-hole symmetry
is present) in a half-filled lattice. It was recently shown in
Ref.~\onlinecite{Oztarhan2025} that realistic artificial graphene
nanostructures with hexagonal armchair geometry exhibit a Nagaoka
ferromagnetic phase stabilized by Coulomb scattering mechanisms in
addition to the Hubbard $U$ term. This phenomenon was overlooked in
previous studies examining the single-electron induced edge
depolarization in TGQDs. To investigate the possibility of Nagaoka
ferromagnetism in artificial TGQDs, Fig. 5 shows the energy difference
between the Nagaoka and edge-depolarized phases,
$E_{Nag}-E_{dep}$, for $N_s=33, 46,$ and $61$ site structures. Across
the investigated parameter range, we observe no signatures of Nagaoka
ferromagnetism. This absence can be attributed to the presence of
zigzag edges, which break electron-hole symmetry and violate the
connectivity condition \cite{Nagaoka1966} required for the Nagaoka
phase, in contrast to hexagonal armchair-edged systems.

In conclusion, we investigated magnetic phases in charge neutral and
gated semiconductor artificial TGQDs up to 61 sites, using continuum
QMC methods, as a function of site the radius $\rho$. For half-filled
lattice, the system remains edge-polarized being consistent with Lieb's
theorem, while going from metallic phase to AFM insulator phase as the
quantum well radius is decreased. If a single electron is added to the system,
spin depolarization of the edge states occurs in both the metallic and
the insulator regimes, validating with previous predictions based on exact
diagonalization results. We have also investigated whether the system
becomes fully polarized due to Nagaoka ferromagnetism. In contrast
with previous recent work on hexagonal armchair artificial graphene
flakes, we have not observed emergence of ferromagnetism as a ground
state when an electron is added, presumably due to presence of zigzag
edges breaking the electron-hole symmetry and violating the
connectivity condition. These observations highlight the dependence of
magnetic correlations on system geometry and finite-size effects, as
revealed through a Hamiltonian model that explicitly include
long-range and exchange interactions, thereby offering insight into
the potential applicability of artificial TGQDs in quantum simulation
and spintronic applications.

\begin{acknowledgments}
  We thank Pawel Potasz for valuable conversations. The quantum Monte Carlo calculations reported in this study were performed using the CHAMP program \cite{Cha-PROG-XX}. This work was supported by The Scientific and
  Technological Research Council of Turkey (TUBITAK) under the 1001 Grant
  Project Number 119F119.  The numerical calculations reported in this study were
  partially performed at TUBITAK ULAKBIM, High Performance and Grid Computing
  site (TRUBA resources).
\end{acknowledgments}

\bibliography{main}

\end{document}